\documentclass{PoS-hep}
\usepackage{amsmath}
\usepackage{bm}
\usepackage{epsfig}
\usepackage{graphics}
\usepackage{upgreek}
\usepackage{amsfonts}
\usepackage{amsbsy}
\usepackage{amscd}
\usepackage{bbm}
\usepackage{eufrak}
\usepackage{cite}

\renewenvironment{subequations}{%
\refstepcounter{equation}%
\setcounter{parentequation}{\value{equation}}%
  \setcounter{equation}{0}
  \ignorespaces
}{%
  \setcounter{equation}{\value{parentequation}}%
  \ignorespacesafterend
}

\newcommand{\Eref}[1]{Eq.~(\ref{#1})}
\newcommand{\Sref}[1]{Sec.~\ref{#1}}
\newcommand{\Fref}[1]{Fig.~\ref{#1}}
\newcommand{\Tref}[1]{Table~\ref{#1}}
\newcommand{\cref}[1]{Ref.~\cite{#1}}

\newcommand{\bal}{\begin{align}}
\newcommand{\eal}{\end{align}}
\newcommand{\beqs}{\begin{subequations}}
\newcommand{\eeqs}{\end{subequations}}
\newcommand{\eec}{\end{center}}
\newcommand{\bec}{\begin{center}}

\newcommand{\eem}{\end{matrix}}
\newcommand{\bem}{\begin{matrix}}
\newcommand{\eeq}{\end{equation}}
\newcommand{\beq}{\begin{equation}}
\newcommand{\ba}{\begin{array}}
\newcommand{\ea}{\end{array}}
\newcommand{\bea}{\begin{eqnarray}}
\newcommand{\eea}{\end{eqnarray}}
\newcommand{\baq}{\begin{eqnarray}}
\newcommand{\eaq}{\end{eqnarray}}

\newcommand\eqs[2]{Eqs.~(\ref{#1}) and (\ref{#2})}

\newcommand{\ftn}{\footnotesize}

\newcommand{\GeV}{{\mbox{\rm GeV}}}

\newcommand{\sEref}[2]{Eq.~(\ref{#1}{\ftn\sf {#2}})}

\newcommand{\etal}{{\it et al.\/}}

\def\to{\rightarrow}

\def\lf{\left(}
\def\rg{\right)}
\newcommand\vev[1]{\langle {#1} \rangle}

\newcommand{\Vhi}{\ensuremath{\widehat V_{\rm CI}}}
\newcommand{\Vjhi}{\ensuremath{V_{\rm CI}}}
\newcommand{\Hhi}{\ensuremath{\widehat H_{\rm CI}}}

\newcommand{\Khi}{\ensuremath{K}}
\newcommand{\Whi}{\ensuremath{W}}
\newcommand{\Vhio}{\ensuremath{\widehat V_{\rm CI0}}}
\newcommand{\Ns}{\ensuremath{{\what N_\star}}}
\newcommand{\ck}{\ensuremath{c_{\rm K}}}

\newcommand{\mP}{\ensuremath{m_{\rm P}}}

\newcommand{\Qef}{\ensuremath{\Lambda_{\rm UV}}}

\def\openone{\leavevmode\hbox{\small1\kern-3.8pt\normalsize1}}
\newcommand{\dV}{\ensuremath{\Delta\widehat V_{\rm CI}}}

\newcommand{\fr}{\ensuremath{f_{ R}}}
\newcommand{\fns}{\ensuremath{f_{n\star}}}
\newcommand{\fms}{\ensuremath{f_{m\star}}}

\newcommand{\fk}{\ensuremath{f_{\rm K}}}
\newcommand{\hk}{\ensuremath{F_{\rm K}}}
\newcommand{\hr}{\ensuremath{F_{ R}}}

\newcommand{\kx}{\ensuremath{k_S}}
\newcommand{\kpp}{\ensuremath{k_\Phi}}
\newcommand{\ksp}{\ensuremath{k_{S\Phi}}}
\newcommand{\ca}{\ensuremath{c_{ R}}}

\newcommand{\ks}{\ensuremath{k_\star}}
\newcommand{\ns}{\ensuremath{n_{\rm s}}}

\newcommand{\as}{\ensuremath{a_{\rm s}}}
\newcommand{\As}{\ensuremath{A_{\rm s}}}
\newcommand{\rw}{\ensuremath{r_{0.002}}}
\newcommand{\rs}{\ensuremath{r_{{R}\rm K}}}
\newcommand{\rcc}{\ensuremath{{R}}}
\newcommand{\rce}{\ensuremath{\widehat{{R}}}}
\newcommand{\Ve}{\ensuremath{\widehat{V}}}

\newcommand{\Ne}{\ensuremath{{\what N}}}

\newcommand{\what}{\ensuremath{\widehat}}

\def\bbet{{\bar\beta}}
\def\al{{\alpha}}

\def\n{\bar{n}}

\def\th{{\theta}}

\newcommand{\sg}{\ensuremath{\phi}}

\newcommand{\sgx}{\ensuremath{\phi_\star}}
\newcommand{\sgf}{\ensuremath{\phi_{\rm f}}}

\newcommand{\ld}{\ensuremath{\lambda}}

\newcommand{\se}{\ensuremath{\widehat \phi}}
\newcommand{\sex}{\ensuremath{\widehat{\phi}_\star}}
\newcommand{\sef}{\ensuremath{\widehat{\phi}_{\rm f}}}
\newcommand{\geu}{\ensuremath{\widehat g}}
\newcommand{\eph}{\ensuremath{\widehat \epsilon}}
\newcommand{\ith}{\ensuremath{\widehat \eta}}

\def\Ka{K\"{a}hler potential}

\def\bcp{{\sc\small Bicep2}/{\it Keck Array}}
\newcommand{\plk}{{\it Planck}}

\title{Observable Gravitational Waves From Kinetically Modified Non-Minimal Inflation}

\ShortTitle{Observable Gravitational Waves From Kinetically
Modified nMI}

\author{\speaker{C. Pallis}\\
Departament de F\'isica Te\`orica and IFIC,\\
Universitat de Val\`encia-CSIC, \\ E-46100 Burjassot, SPAIN}

\abstract{We consider Supersymmetric (SUSY) and non-SUSY models of
chaotic inflation based on the simplest power-law potential with
exponents $n=2$ and $4$. We propose a convenient non-minimal
coupling to gravity and a non-minimal kinetic term which ensure,
mainly for $n=4$, inflationary observables favored by the \bcp\
and \plk\ results. Inflation can be attained for subplanckian
inflaton values with the corresponding effective theories
retaining the perturbative unitarity up to the Planck scale.
\\ \\ {\sl\bfseries
Published in}~~{PoS  PLANCK {\bf 2015}, 095 (2015)}. }

\FullConference{18th International Conference From the Planck Scale to the Electroweak Scale \\
                  25-29 May 2015\\
                  Ioannina, Greece}

\begin{document}

\section{Introduction}

Kinetically modified \emph{Non-minimal} (chaotic) \emph{inflation}
({\sf\small nMI}) \cite{nMkin} is a variant of nMI which arises in
the presence of a non-canonical kinetic term for the inflaton
$\sg$ -- apart from the non-minimal coupling $\fr(\sg)$ between
$\sg$ and the Ricci scalar curvature, $\rcc$ which is required by
definition in any model of nMI \cite{old}. In this talk we focus
on inflationary models based on a synergy between $\fr$ and the
inflaton potential $\Vjhi$, which are selected  \cite{nMkin,nmi,
roest} as follows
\beq
\Vjhi(\sg)=\ld^2\sg^n/2^{n/2}\>\>\>\>\>\>\mbox{and}\>\>\>\>\>\>
\fr=1+\ca\sg^{n/2}\>\>\>\>\>\>\mbox{with}\>\>\>\>\>\>n=2,4\,.\label{Vn}
\eeq
Below, we first (in Sec.~\ref{Fhi}) briefly review the basic
ingredients of nMI in a non-\emph{Supersymmetric} ({\sf\small
SUSY}) framework and constrain the parameters of the models in
Sec.~\ref{res0} taking into account a number of observational and
theoretical requirements described in \Sref{obs}. Then (in
Sec.~\ref{Qef}) we focus on the problem with perturbative
unitarity that plagues \cite{cutoff, riotto} these models  at the
strong coupling and motivate the form of $\fk$ analyzed in our
work.

Throughout the text, the subscript $,\chi$ denotes derivation
\emph{with respect to} ({\sf\ftn w.r.t}) the field $\chi$, charge
conjugation is denoted by a star ($^*$) and we use units where the
reduced Planck scale $\mP = 2.43\cdot 10^{18}~\GeV$ is set equal
to unity.

\subsection{Coupling non-Minimally the Inflaton to Gravity}\label{Fhi}

The action of the inflaton $\sg$ in the \emph{Jordan frame}
({\sf\ftn JF}), takes the form:
\beq \label{action1} {\sf  S} = \int d^4 x \sqrt{-\mathfrak{g}}
\left(-\frac{\fr}{2}\rcc +\frac{\fk}{2}g^{\mu\nu}
\partial_\mu \sg\partial_\nu \sg-
\Vjhi(\sg)\right). \eeq
where $\mathfrak{g}$ is the determinant of the background
Friedmann-Robertson-Walker metric, $g^{\mu\nu}$ with signature
$(+,-,-,-)$, $\vev{\fr}\simeq1$ to guarantee the ordinary Einstein
gravity at low energy and we allow for a kinetic mixing through
the function $\fk(\phi)$. By performing a conformal transformation
\cite{nmi} according to which we define the \emph{Einstein frame}
({\sf\ftn EF}) metric $\geu_{\mu\nu}=\fr\,g_{\mu\nu}$ we can write
${\sf S}$ in the EF as follows
\beqs\beq {\sf  S}= \int d^4 x
\sqrt{-\what{\mathfrak{g}}}\left(-\frac12
\rce+\frac12\geu^{\mu\nu} \partial_\mu \se\partial_\nu \se
-\Vhi(\se)\right), \label{action} \eeq
where hat is used to denote quantities defined in the EF. We also
introduce the EF canonically normalized field, $\se$, and
potential, $\Vhi$, defined as follows:
\beq \label{VJe}
\frac{d\se}{d\sg}=J=\sqrt{\frac{\fk}{\fr}+{3\over2}\left({f_{
R,\sg}\over \fr}\right)^2}\>\>\>~\mbox{and}\>\>\>~ \Vhi=
\frac{\Vjhi}{\fr^2}\,,\eeq\eeqs
where the symbol $,\phi$ as subscript denotes derivation w.r.t the
field $\phi$. Plugging \Eref{Vn} into \Eref{VJe}, we obtain
\beq \label{Vhio}
J^2=\frac{\fk}{\fr}+\frac{3n^2\ca^2\sg^{n-2}}{8\fr^2}\>\>\>
\>\>\>~ ~\mbox{and}\>\>\>\>\>\>~
\Vhi=\frac{\ld^2\sg^n}{2^{n/2}\fr^{2}}.\eeq
In the pure nMI \cite{old, nmi, roest} we take $\fk=1$ and, for
$\ca\gg1$, we infer from \Eref{VJe}, that $\fr$ determines the
relation between $\se$ and $\sg$ and controls the shape of $\Vhi$
affecting thereby the observational predictions -- see below.

\subsection{Inflationary Observables -- Constraints} \label{obs}

A model of nMI can be qualified as successful, if it can become
compatible with the following observational and theoretical
requirements:

\paragraph{\bf 1.2.1} The number of e-foldings $\Ns$ that the scale $\ks=0.05/{\rm Mpc}$
experiences during this nMI must to be enough for the resolution
of the horizon and flatness problems of standard Big Bang, i.e.,
\cite{plcp}
\begin{equation}
\label{Nhi}  \Ns=\int_{\sef}^{\sex} d\se\frac{\Vhi}{\Ve_{\rm
CI,\se}}\simeq55,\eeq
where $\sgx~[\sex]$ are the value of $\sg~[\se]$ when $\ks$
crosses the inflationary horizon. Also $\sg_{\rm f}~[\se_{\rm f}]$
is the value of $\sg~[\se]$ at the end of nMI, which can be found,
in the slow-roll approximation, from the condition
$$ {\ftn\sf max}\{\widehat\epsilon(\sg_{\rm
f}),|\widehat\eta(\sg_{\rm f})|\}=1,\>\>\>~\mbox{where}$$
\beq \label{sr}\widehat\epsilon= {1\over2}\left(\frac{\Ve_{\rm
CI,\se}}{\Ve_{\rm CI}}\right)^2={1\over2J^2}\left(\frac{\Ve_{\rm
CI,\sg}}{\Ve_{\rm CI}}\right)^2
\>\>\>\mbox{and}\>\>\>\widehat\eta= \frac{\Ve_{\rm
CI,\se\se}}{\Ve_{\rm CI}}={1\over J^2}\left(\frac{\Ve_{\rm
CI,\sg\sg}}{\Ve_{\rm CI}}-\frac{\Ve_{\rm CI,\sg}}{\Ve_{\rm
CI}}{J_{,\sg}\over J}\right)\cdot \eeq
It is evident from the formulas above that non trivial
modifications on $\fk$ and thus to $J$ may have an pronounced
impact on the parameters above modifying thereby the inflationary
observables too.

\paragraph{\bf 1.2.2} The amplitude $\As$ of the power spectrum of the curvature perturbation
generated by $\sg$ at  $k_{\star}$ has to be consistent with
data~\cite{plcp}, i.e.,
\begin{equation}  \label{Prob}
\sqrt{A_{\rm s}}=\: \frac{1}{2\sqrt{3}\, \pi} \; \frac{\Ve_{\rm
CI}(\sex)^{3/2}}{|\Ve_{\rm
CI,\se}(\sex)|}={1\over2\pi}\,\sqrt{\frac{\Vhi(\sgx)}{6\what\epsilon_\star}}
\simeq4.627\cdot 10^{-5},
\end{equation}
where the variables with subscript $\star$ are evaluated at
$\sg=\sgx$.

\paragraph{\bf 1.2.3} The remaining inflationary observables (the spectral index $\ns$,
its running $\as$, and the tensor-to-scalar ratio $r$) --
estimated through the relations:
\beq\label{ns} \mbox{\ftn\sf (a)}\>\>\ns=\: 1-6\eph_\star\ +\
2\ith_\star,\>\>\>~\mbox{\ftn\sf (b)}\>\> \as
=\frac23\left(4\ith^2-(\ns-1)^2\right)-2\what\xi_\star\>\>\>~
\mbox{and}\>\>\>~\mbox{\ftn\sf (c)}\>\>r=16\eph_\star\,, \eeq
with $\widehat\xi={\Ve_{\rm CI,\se} \Ve_{\rm
CI,\se\se\se}/\Ve_{\rm CI}^2}$ -- have to be consistent with the
data \cite{plcp},  i.e.,
\begin{equation}  \label{nswmap}
\mbox{\ftn\sf
(a)}\>\>\ns=0.968\pm0.009\>\>\>~\mbox{and}\>\>\>~\mbox{\ftn\sf
(b)}\>\>r\leq0.12,
\end{equation}
at 95$\%$ \emph{confidence level} ({\sf\ftn c.l.}) -- pertaining
to the $\Lambda$CDM$+r$ framework with $|\as|\ll0.01$. Although
compatible with \sEref{nswmap}{b} the present combined \plk\ and
\bcp\ results \cite{gws} seem to favor $r$'s of order $0.01$ since
$r= 0.048^{+0.035}_{-0.032}$ at 68$\%$ c.l. has been reported.

\paragraph{\bf 1.2.4} The effective theory describing nMI has to remains
valid up to a UV cutoff scale $\Qef$ to ensure the stability of
our inflationary solutions, i.e.,
\beq \label{subP}\mbox{\ftn\sf (a)}\>\> \Vhi(\sgx)^{1/4}\leq\Qef
\>\>\>~\mbox{and}\>\>\>~\mbox{\ftn\sf (b)}\>\>\sgx\leq\Qef.\eeq
It is expected that $\Qef\simeq\mP$, contrary to the pure nMI with
$\ca\gg1$ where $\Qef\ll\mP$ -- see \Sref{Qef}.

\subsection{The Two Regimes of Synergistic nMI} \label{res0}

The models of nMI based on \Eref{Vn} exhibit the following two
regimes:

\paragraph{\bf 1.3.1} The \emph{weak} $\ca$ regime with $\ca\leq1$. In this case
from \Eref{VJe} we find $J\simeq1/\fr$ and applying \eqs{Nhi}{sr},
the slow-roll parameters and $\Ns$ read
\beq\label{nmci2b} \eph\simeq\frac{n^2}{2 \sg^2
\fr},\>\>\>\ith\simeq2\lf1-\frac1n\rg\eph-\frac{4+n}{2n}
\ca\sg^\frac{n}{2}\eph\,
\>\>\>\>\>\>\mbox{and}\>\>\>\>\>\>\Ns\simeq\frac{\sgx^2}{2n}\cdot\eeq
Imposing the condition of \Eref{sr} and solving then the latter
equation w.r.t $\sg_*$ we arrive at
\beq\label{nmci4b}\sgf\simeq
n/\sqrt{2}\>\>\>\>\>\>\mbox{and}\>\>\>\>\>\>
\sgx\simeq\sqrt{2n\Ns}\,.\eeq
Inflation is attained, thus, only for $\sg>1$. On the other hand,
\Eref{Prob} implies
\beq \label{Prob2}\ld = \sqrt{6\As\fns}\pi
n^{(2-n)/4}/\Ns^{(2+n)/4}\,,\eeq
where $\fns=\fr(\sgx)=1 +\ca(2n\Ns)^{n/4}$. Applying \Eref{ns} we
find that the inflationary observables are $\ca$-dependent and can
be marginally consistent with \Eref{nswmap} -- see \Sref{res2}.
Indeed,
\beqs\bea\label{ns1}&& \ns=1 - \lf4 + n +n/\fns\rg/4\Ns,\>\>\>r=4
n/\fns\Ns,\>\>\>~~~~~~~~~~\\ && \hspace*{-2.5mm}\as=
\big(n^2-n(n+4)\fns-4(n+4)\fns^2\big)/16\fns^2\Ns^2\,.~~~
\label{as1}\eea\eeqs
In the limit $\ca\to0$ or $\fns\to1$ the results of the simplest
power-law chaotic inflation -- with $\fr=\fk=1$ and $V_{\rm CI}$
given in \Eref{Vn} -- are recovered. These are by now disfavored
by \Eref{nswmap}.

\paragraph{\bf 1.3.2} The \emph{strong} $\ca$ regime with $\ca\gg1$. In this case, from
\Eref{VJe} we find
\beq
J\simeq{\sqrt{3}n\ca\sg^{n/2-1}/2\sqrt{2}\fr}\>\>\>\>\>\>\mbox{and}\>\>\>\>\>\>\Vhi\simeq
\ld^2/2^{n/2}\ca^2.\>\>\eeq
We observe that $\Vhi$ exhibits an almost flat plateau. From
\eqs{Nhi}{sr} we find
\beq\eph\simeq {4/3\ca^2\sg^n},\>\>\>
\ith\simeq-{4/3\ca\sg^{n/2}}\>\>\>\>\>\>\mbox{and}\>\>\>\>\>\>\Ns\simeq{3\ca\sg_*^{n/2}/4}\,.\eeq
Therefore, $\sgf$ and $\sgx$ are found from the condition of
\Eref{sr} and the last equality above, as follows
\beq \label{sigs1}\sgf=\mbox{\sf\ftn
max}\{(4/3\ca^2)^{1/n},(4/3\ca)^{2/n}\}
\>\>\>\>\>\>\mbox{and}\>\>\>\>\>\> \sgx=(4\Ns/3\ca)^{2/n}.\eeq
Consequently, nMI can be achieved even with subplanckian $\sg$
values for $\ca\gtrsim (4\Ne_*/3)^{2/n}$. Also the normalization
of \Eref{Prob} implies the following relation between $\ca$ and
$\ld$
\beq \label{Prob1} A^{1/2}_{\rm s}\simeq
\left.2^{-(10+n)/4}\frac{\ld\ca\sg^n}{\pi\fr}\right|_{\sg=\sgx}\>\>\>\Rightarrow\>\>\>
\ld\simeq\frac{3\cdot2^{n/4}}{\Ns}\sqrt{2\As}\;\pi\;\ca\,.\eeq
From \Eref{ns} we obtain the $\ca$-independent values for the
observables:
\beq \label{nswmap1} \ns\simeq1-{2/\Ns}\simeq0.965, \>\>\> \as
\simeq{-2/\Ns^2}\simeq-6.4\cdot10^{-4}\>\>\>\>\>\>
\mbox{and}\>\>\>\>\>\> r\simeq{12/\Ns^2}\simeq4\cdot 10^{-3},\eeq
which are in agreement with \Eref{nswmap}, although with low
enough $r$ values.

\subsection{The Ultraviolet (UV) Cut-off Scale} \label{Qef}

In the highly predictive regime with large $\ck$, the models
violate {perturbative unitarity} for $n>2$. To see this we analyze
the small-field behavior of the theory in order to extract the UV
cut-off scale $\Qef$. The result depends crucially on the value of
$J$ in \Eref{VJe} in the vacuum, $\vev{\phi}=0$. Namely we have
\beq
\vev{J}=\begin{cases}\sqrt{3/2}\ca\>\>&\mbox{for}\>\>n=2,\\1\>\>&\mbox{for}\>\>n\neq2\,.\end{cases}\eeq
For $n=2$ and any $\ca$ we obtain $\se\neq\sg$. Expanding the
second and third term of ${\sf S}$ in the right-hand side of
\Eref{action} about $\vev{\phi}=0$ in terms of $\se$ we obtain:
\beq J^2
\dot\phi^2=\lf1-\sqrt{\frac{8}{3}}{\se}+2{\se^2}-\cdots\rg\dot\se^2
\>\>\>\>\>\>\mbox{and}\>\>\>\>\>\>\Vhi=\frac{\ld^2\what
\phi^2}{3\ca^2}\lf1-\sqrt{\frac{8}{3}}\se+ 2\se^2-\cdots\rg.\eeq
As a consequence $\Qef=\mP$ since the expansions above are $\ca$
independent. On the contrary, for $n>2$ we have $\se=\sg$ and the
expansions of the same terms in \Eref{action} are $\ca$ dependent:
\beqs\bea && J^2
\dot\phi^2=\lf1-\ca\what{\sg}^\frac{n}{2}+\frac{3n^2}{8}\ca^2\what{\sg}^{n-2}+
\ca^2\what{\sg}^{n}-\cdots\rg\dot\se^2; \label{Jexp}\\
&&
\Vhi=\frac{\ld^2\what{\sg}^n}{2}\lf1-2\ca\what{\sg}^{\frac{n}{2}}+3\ca^2\what{\sg}^n-
4\ca^3\what{\sg}^{\frac{3n}{2}}+\cdots\rg\cdot \label{Vexp}
\eea\eeqs
Since the term which yields the smallest denominator for $\ca>1$
is  ${3n^2}\ca^2\se^{n-2}/8$ we find \cite{cutoff,riotto}:
\beq \Qef=\mP/\ca^{2/(n-2)}\ll\mP\,. \label{Luv} \eeq

However, if we introduce a \emph{non-canonical kinetic mixing} of
the form
\beq\label{fk}
\fk(\sg)=\ck\fr^m\>\>\>\>\>\>\mbox{where}\>\>\>\>\>\>
\ck=(\ca/\rs)^{4/n}\>\>\>\>\>\>\mbox{and}\>\>\>\>\>\>m\geq0,\eeq
no problem with the perturbative unitarity emerges for $\rs\leq1$,
even if $\ca$ and/or $\ck$ are large -- the latter situation is
expected if we wish to achieve efficient nMI with $\sg\leq1$.
E.g., for $m=0$ the expansions in \eqs{Jexp}{Vexp} can be
rewritten replacing $\ca$ with $\rs$ and $\ld$ with
$\ld/\ck^{n/4}$ -- similar expressions can be obtained for other
$m$, too. In other words, the perturbative unitarity can be
preserved up to $\mP$ if we select a non-trivial $\fk$ such that
$\vev{J}\neq1$. This requirement lets a functional uncertainty as
regards the form of $\fk$ during nMI which can be parameterized as
shown in \Eref{fk} given that $\vev{\fr}\simeq1$ -- see
\Sref{Fhi}.

We below describe a possible formulation of this type of nMI in
the context of \emph{Supergravity} ({\sf\ftn SUGRA}) -- see
\Sref{sugra} -- and then, in \Sref{res}, we analyze the
inflationary behavior of these models. We conclude summarizing our
results in \Sref{con}.

\section{Supergravity Embeddings} \label{sugra}

The models above -- defined by \eqs{Vn}{fk} -- can be embedded in
SUGRA if we use two gauge singlet chiral superfields $z^\al=\Phi,
S$, with $\Phi$ ($\al=1$) and $S$ ($\al=2)$ being the inflaton and
a ``stabilizer'' field respectively. The EF action for $z^\al$'s
can be written as \cite{linde1}
\beqs \beq\label{Saction1}  {\sf S}=\int d^4x \sqrt{-\what{
\mathfrak{g}}}\lf-\frac{1}{2}\rce \ +\ K_{\al\bbet}
\geu^{\mu\nu}\partial_\mu z^\al \partial_\nu z^{*\bbet}-\Ve\rg\,,
\eeq
where summation is taken over the scalar fields $z^\al$, $\Khi$ is
the \Ka\ with $K_{\al\bbet}=K_{,z^\al z^{*\bbet}}$ and
$K^{\al\bbet}K_{\bbet\gamma}=\delta^\al_{\gamma}$. Also $\Ve$ is
the EF F--term SUGRA potential given by
\beq \Ve=e^{\Khi}\left(K^{\al\bbet}D_\al W D^*_\bbet W^*-3{\vert
W\vert^2}\right),\label{Vsugra} \eeq \eeqs
where $D_\al W=W_{,z^\al} +K_{,z^\al}W$ with $\Whi$ being the
superpotential. Along the inflationary track determined by the
constraints
\beq \label{inftr} S=\Phi-\Phi^*=0,\>\>\>\mbox{or}\>\>\>s=\bar
s=\th=0\eeq
\begin{floatingtable}[r]
\begin{tabular}{|l||ll|}\hline
{\sc Superfields}&$S$&$\Phi$\\\hline\hline
$U(1)_R$ &$1$&$0$\\
$U(1)$&$-1$&$2/n$\\\hline
\end{tabular}
\caption {\sl Charge assignments of the superfields.}\label{ch}
\end{floatingtable}
\noindent if we express $\Phi$ and $S$ according to the
parametrization
\beq \Phi=\:{\phi\,e^{i
\th}}/{\sqrt{2}}\>\>\>\>\>\>\mbox{and}\>\>\>\>\>\>S=\:(s +i\bar
s)/\sqrt{2}\,,\label{cannor} \eeq
$V_{\rm CI}$ in \Eref{Vn} can be produced, in the flat limit, by
\beq \label{Wn} W=\ld S\Phi^{n/2}.\eeq
The form of $W$ can be uniquely determined if we impose an $R$ and
a global $U(1)$ symmetry with charge assignments shown in
\Tref{ch}.


On the other hand, the derivation of $\Vhi$ in \Eref{Vhio} via
\Eref{Vsugra} requires a judiciously chosen $K$. Namely, along the
track in \Eref{inftr} the only surviving term in \Eref{Vsugra} is
\beq \label{1Vhio}\Vhi=\Ve(\th=s=\bar s=0)=e^{K}K^{SS^*}\,
|W_{,S}|^2\,.\eeq
The incorporation $\fr$ in \Eref{Vn} and $\fk$ in \Eref{fk}
dictates the adoption of a logarithmic $K$ \cite{linde1} including
the functions
\beq \label{hr}
\hr(\Phi)=1+2^{\frac{n}{4}}\Phi^{\frac{n}{2}}\ca\,,\>\>\>\hk=(\Phi-\Phi^*)^2\,\>\>\>\>\>\>\mbox{and}\>\>\>\>\>\>F_S={|S|^2}-\kx{|S|^4}\,.
\eeq
Here, \hr\ is an holomorphic function reducing to $\fr$, along the
path in \Eref{inftr}, $\hk$ is a real function which assists us to
incorporate the non-canonical kinetic mixing generating by $\fk$
in \Eref{fk}, and $F_S$ provides a typical kinetic term for $S$,
considering the next-to-minimal term for stability/heaviness
reasons \cite{linde1}. Indeed, $\hk$ lets intact $\Vhi$, since it
vanishes along the trajectory  in \Eref{inftr}, but it contributes
to the normalization of $\Phi$. Taking for consistency all the
possible terms up to fourth order, $K$ is written as
\beqs\beq
K_1=-3\ln\left(\frac12\lf\hr+\hr^*\rg+\frac{\ck}{3\cdot2^{m+1}}\lf\hr+
\hr^{*}\rg^{m}\hk-\frac13F_S+\frac\kpp6 \hk^2-\frac\ksp3
\hk{|S|^2}\right)\,.~~~~~\label{K1}\eeq
Alternatively, if we do not insist on a pure logarithmic $K$, we
could also adopt the form
\beq
K_2=-3\ln\left(\frac12\lf\hr+\hr^*\rg-\frac13F_S\right)-\frac{\ck}{2^m}\frac{\hk}{\lf\hr+
\hr^{*}\rg^{1-m}}\,\cdot\label{K2}\eeq
Moreover, if we place $F_S$ outside the argument of the logarithm
similar results are obtained by the following $K$'s -- not
mentioned in \cref{nMkin}:
\bea \label{K3} &&
K_3=-2\ln\left(\frac12\lf\hr+\hr^*\rg+\frac{\ck}{2^{m+2}}\lf\hr+
\hr^{*}\rg^{m}\hk \right)+F_S\,,\\ &&
K_4=-2\ln\frac{\hr+\hr^*}{2}-\frac{\ck}{2^m}\frac{\hk}{\lf\hr+
\hr^{*}\rg^{1-m}}+F_S\,.\label{K4}\eea\eeqs
Note that for $m=0$ [$m=1$], $\hr$ and $\hk$ in $K_1$ and $K_3$
[$K_2$ and $K_4$] are totally decoupled, i.e. no higher order term
is needed. Also we use only integer prefactors for the logarithms
avoiding thereby any relevant tuning -- cf.~\cref{nIG}. Our
models, for $\ck\gg\ca$, are completely natural in the 't Hooft
sense because, in the limits $\ca\to0$ and $\ld\to0$, the theory
enjoys the enhanced symmetries
\beq \Phi \to\ \Phi^*,\> \Phi \to\ \Phi+c
\>\>\>\>\>\>\mbox{and}\>\>\>\>\>\> S \to\ e^{i\alpha}
S,\label{shift}\eeq
where $c$ is a real number. It is evident that our proposal is
realized more attractively within SUGRA than within the non-SUSY
set-up, since both $\fk$ and $\fr$ originate from the same
function $K$.

To verify the appropriateness of $K$'s in Eqs.~(\ref{K1}) --
(\ref{K4}), we can first remark that, along the trough in
\Eref{inftr}, these are diagonal with non-vanishing elements
$K_{SS^*}$ and $K_{\Phi\Phi^*}=J^2$, where $J$ is given by
\Eref{Vhio} for $K=K_i$ and \Eref{Vhio} replacing $3/8$ by $1/4$
for $K=K_{i+2}$. Substituting into \Eref{1Vhio}
$K^{SS^*}=1/K_{SS^*}$ and $\exp K=1/\fr^{N}$, where
\beq \label{Nab} K_{SS^*}=\begin{cases}
1/\fr\\1\end{cases}\mbox{and}\>\>\>\>N=\begin{cases}
3\\2\end{cases} \mbox{for}\>\>\>\>\>\>K=\begin{cases} K_i\\
K_{i+2} \end{cases}\mbox{with}~~i=1,2, \eeq
we easily deduce that $\Vhi$ in \Eref{Vhio} is recovered. If we
perform the inverse of the conformal transformation described in
\eqs{action}{action1} with frame function
${\Omega/N}=-e^{-{K}/{N}}$  we can easily show that
$\fr=-\Omega/N$ along the path in \Eref{inftr}. Note, finally,
that the conventional Einstein gravity is recovered at the SUSY
vacuum, $\vev{S}=\vev{\Phi}=0$, since $\vev{\fr}\simeq1$.

\renewcommand{\arraystretch}{1.2}
\begin{table*}[t]
\bec
\begin{tabular}{|c|c|c|c|c|c|}\hline
{\sc Fields}&{\sc Eingestates} & \multicolumn{4}{c|}{\sc Masses
Squared}\\\cline{3-6}
&&{\sc Symbol} & {$K=K_1$}&{$K=K_2$} &{$K=K_{i+2}$} \\
\hline
2 real scalars&$\widehat\theta$&$\widehat m_{\theta}^2$& $4\Hhi^2$
&\multicolumn{2}{|c|}{$6\Hhi^2$}\\\cline{3-6}
1 complex scalar&$\widehat s, \widehat{\bar{s}}$ & $ \widehat m_{
s}^2$&\multicolumn{2}{c|}{$6\lf2\kx\fr-1/3\rg\Hhi^2$}&$12\kx\Hhi^2$\\\hline
$4$ Weyl spinors & $\what \psi_\pm $ & $\what m^2_{ \psi\pm}$ &
\multicolumn{3}{c|}{$3n^2 \Hhi^2/2\ck\sg^2\fr^{1+m}$}
\\\hline
\end{tabular}\label{tab1}\eec
\caption{\normalfont Mass-squared spectrum for $K=K_i$ and
$K=K_{i+2}$ ($i=1,2$) along the path in Eq.~(2.2).}
\end{table*}

Defining the canonically normalized fields via the relations
${d\widehat \sg/ d\sg}=\sqrt{K_{\Phi\Phi^*}}=J$, $\what{\th}=
J\th\sg$ and $(\what s,\what{\bar s})=\sqrt{K_{SS^*}} {(s,\bar
s)}$ we can verify that the configuration in \Eref{inftr} is
stable w.r.t the excitations of the non-inflaton fields. Taking
the limit $\ck\gg\ca$ we find the expressions of the masses
squared $\what m^2_{\chi^\al}$ (with $\chi^\al=\th$ and $s$)
arranged in \Tref{tab1}, which approach rather well the quite
lengthy, exact formulas. From these expressions we appreciate the
role of $\kx>0$ in retaining positive $\what m^2_{s}$. Also we
confirm that $\what m^2_{\chi^\al}\gg\Hhi^2=\Vhio/3$ for
$\sgf\leq\sg\leq\sgx$. In \Tref{tab1} we display the masses $\what
m^2_{\psi^\pm}$ of the corresponding fermions too with eignestates
$\what \psi_\pm=(\what{\psi}_{\Phi}\pm \what{\psi}_{S})/\sqrt{2}$,
defined in terms of $\what\psi_{S}=\sqrt{K_{SS^*}}\psi_{S}$ and
$\what\psi_{\Phi}=\sqrt{K_{\Phi\Phi^*}}\psi_{\Phi}$, where
$\psi_\Phi$ and $\psi_S$ are the Weyl spinors associated with $S$
and $\Phi$ respectively. Note, finally, that $\what
m_{\chi^\al}\ll\mP$, for any $\chi^\al$, contrary to similar cases
\cite{lazarides} where the inflaton belongs to gauge non-singlet
superfields.

Inserting the derived mass spectrum in the well-known
Coleman-Weinberg formula, we can find the one-loop radiative
corrections, $\dV$ to $\Vhi$. It can be verified that our results
are immune from $\dV$, provided that the renormalization group
mass scale $\Lambda$, is determined conveniently and $\ksp$ and
$\kx$ are confined to values of order unity.

\section{Results}\label{res}

The present inflationary scenario depends on the parameters: $ n,~
m, ~\rs,~ \ld/\ck^{n/4} $. Note that the two last combinations of
parameters above replace $\ck$, $\ca$ and $\ld$. This is because,
if we perform a rescaling $\sg=\tilde\sg/\sqrt{\ck}$,
\Eref{action1} preserves its form replacing $\sg$ with $\tilde\sg$
and $\fk$ with $\fr^m$ where $\fr$ and $\Vjhi$ take, respectively,
the forms
\beq\label{frVrs}
\fr=1+\rs\tilde\sg^{n/2}\>\>\>\>\>\>\mbox{and}\>\>\>\>\>\>
\Vjhi=\ld^2\tilde\sg^{n}/2^{n/2}\ck^{n/2},\eeq
which, indeed, depend only on $\rs$ and $\ld^2/\ck^{n/2}$.
Imposing the restrictions of \Sref{obs} we can delineate the
allowed region of these parameters. Below we first extract some
analytic expressions -- see \Sref{res1} -- which assist us to
interpret the exact numerical results presented in \Sref{res2}.

\subsection{Analytic Results}\label{res1}

Assuming $\ck\gg\ca$, \Eref{VJe} yields
$J\simeq\sqrt{\ck}/\fr^{(1-m)/2}$. Inserting the last one and
$\Vhi$ from \Eref{Vn} in \Eref{sr} we extract the slow-roll
parameters for this model as follows -- cf. \Eref{nmci2b}:
\beq\label{sr1}\eph={n^2}/{2 \sg^2\ck
\fr^{1+m}}\>\>\>\>\>\>\mbox{and}\>\>\>\>\>\>
{\ith}=2\lf1-1/n\rg\eph-\lf{4+n(1+m)}\rg\ca\sg^{n/2}\eph/{2n}
\,.\eeq
Given that $\sg\ll1$ and so $\fr\simeq1$, nMI terminates for
$\sg=\sgf$ found by the condition
\beq \sgf\simeq\mbox{\ftn\sf
max}\{{n/\sqrt{2\ck}},\sqrt{(n-1)n/\ck}\}\,, \label{sgf}\eeq
in accordance with \Eref{sr}. Since $\sgx\gg\sgf$, from \Eref{Nhi}
we find
\begin{equation}
\label{Nhia} \Ns=\frac{\ck\sgx^2}{2n}
\;{}_2F_1\lf-m,4/n;1+4/n;-\ca\sgx^{n/2}\rg=\begin{cases}{\ck\sgx^2}/{2n}\>\>\>&\mbox{for}\>\>\>m=0,\\
(\fr^{1+m}-1)/8(1+m)\rs\>\>\>&\mbox{for}\>\>\>n=4,
\end{cases}
\end{equation}
where ${}_2F_1$ is the Gauss hypergeometric function.
Concentrating on the cases with $m=0$ or $n=4$, we solve
\Eref{Nhia} w.r.t $\sgx$ with results
\begin{equation}
\label{sgx}\sgx\simeq\begin{cases}\sqrt{2n\Ns/\ck}\>\>\>&\mbox{for}\>\>\>\>\>\>m=0,\\
\sqrt{\fms-1}/\sqrt{\rs\ck}\>\>\>&\mbox{for}\>\>\>\>\>\>n=4,
\end{cases}
\end{equation}
where $\fms^{1+m}=1+8(m+1)\rs\Ns$. In both cases there is a lower
bound on $\ck$, above which $\sgx<1$ and so, our proposal can be
stabilized against corrections from higher order terms -- e.g.,
for $n=4, m=1$ and $\rs=0.03$ we obtain
$140\lesssim\ck\lesssim1.4\cdot10^6$ for
$3.3\cdot10^{-4}\lesssim\ld\lesssim3.5$. The correlation between
$\ld$ and $\ck^{n/4}$ can be found from \Eref{Prob}. For $m=0$
this is given by \Eref{Prob2} replacing $\ld$ with $\ld/\ck^{n/4}$
and $\ca$ with $\rs$ in the definition of $\fns$. For $n=4$ we
obtain
\beq \label{lan}\ld = 16\sqrt{3\As}\,\pi\,
\ck\,\rs^{3/2}/(\fms-1)^{3/2}\fms^{(1+m)/2}\,.\eeq

As regards the inflationary observables, these are obviously given
by \eqs{ns1}{as1} for the trivial case with $m=0$. For $m\neq0$,
however, these are heavily altered. In particular, for $n=4$ we
obtain
\beqs\bea\label{ns2}&& \ns=1 - 8\rs\frac{
m-1-(m+2)\fms}{(\fms-1)\fms^{1+m}},\>\>\>r=\frac{128\rs}{(\fms-1)\fms^{1+m}},
\\&& \>\>\>\as=\frac{64\rs^2(1+m)(m+2)}{(\fms-1)^2\fms^{4(1+m)}}\fms^2\lf\fms^{2m}\lf\frac{1-m}{m+2}+\frac{2m-1}{m+1}\fms\rg-\fms^{2(1+m)}\rg.~~~
\label{as2}\eea\eeqs
The formulae above is valid only for $\rs>0$ -- see \Eref{sgx} --
and is simplified \cite{nMkin} for low $m$'s.


\subsection{Numerical Results}\label{res2}

The conclusions obtained in \Sref{res1} can be verified and
extended to others $n$'s and $m$'s numerically. In particular,
enforcing \eqs{Nhi}{Prob} we can restrict $\sgx$ and
$\ld/\ck^{n/4}$. Then we can compute the model predictions via
\Eref{ns}, for any selected $m, n$ and $\rs$. The outputs, encoded
as lines in the $\ns-\rw$ plane, are compared against the
observational data \cite{plcp,gws} in \Fref{fig1} for $n=2$ (left
panel) and $4$ (right panel) setting $m=0,1$ and $4$ -- dashed,
solid, and dot-dashed lines respectively. The variation of $\rs$
is shown along each line. To obtain an accurate comparison, we
compute $\rw=16\eph(\sg_{0.002})$ where $\sg_{0.002}$ is the value
of $\sg$ when the scale $k=0.002/{\rm Mpc}$, which undergoes
$\what N_{0.002}=(\Ns+3.22)$ e-foldings during nMI, crosses the
inflationary horizon.

\begin{figure}[!t]\vspace*{-.12in}
\hspace*{-.19in}
\begin{minipage}{8in}
\epsfig{file=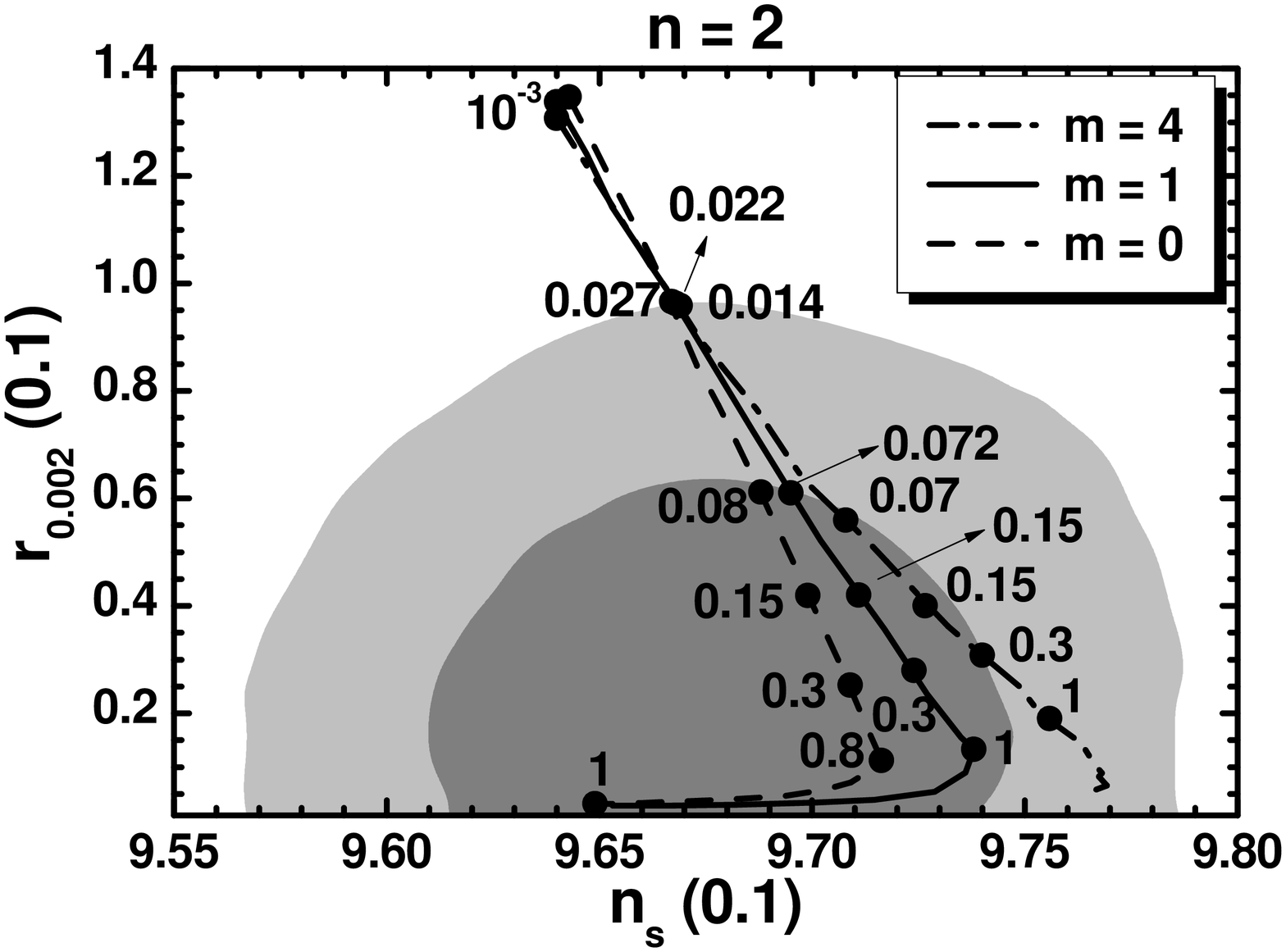,height=3.48in,angle=-90}
\hspace*{-1.2cm}
\epsfig{file=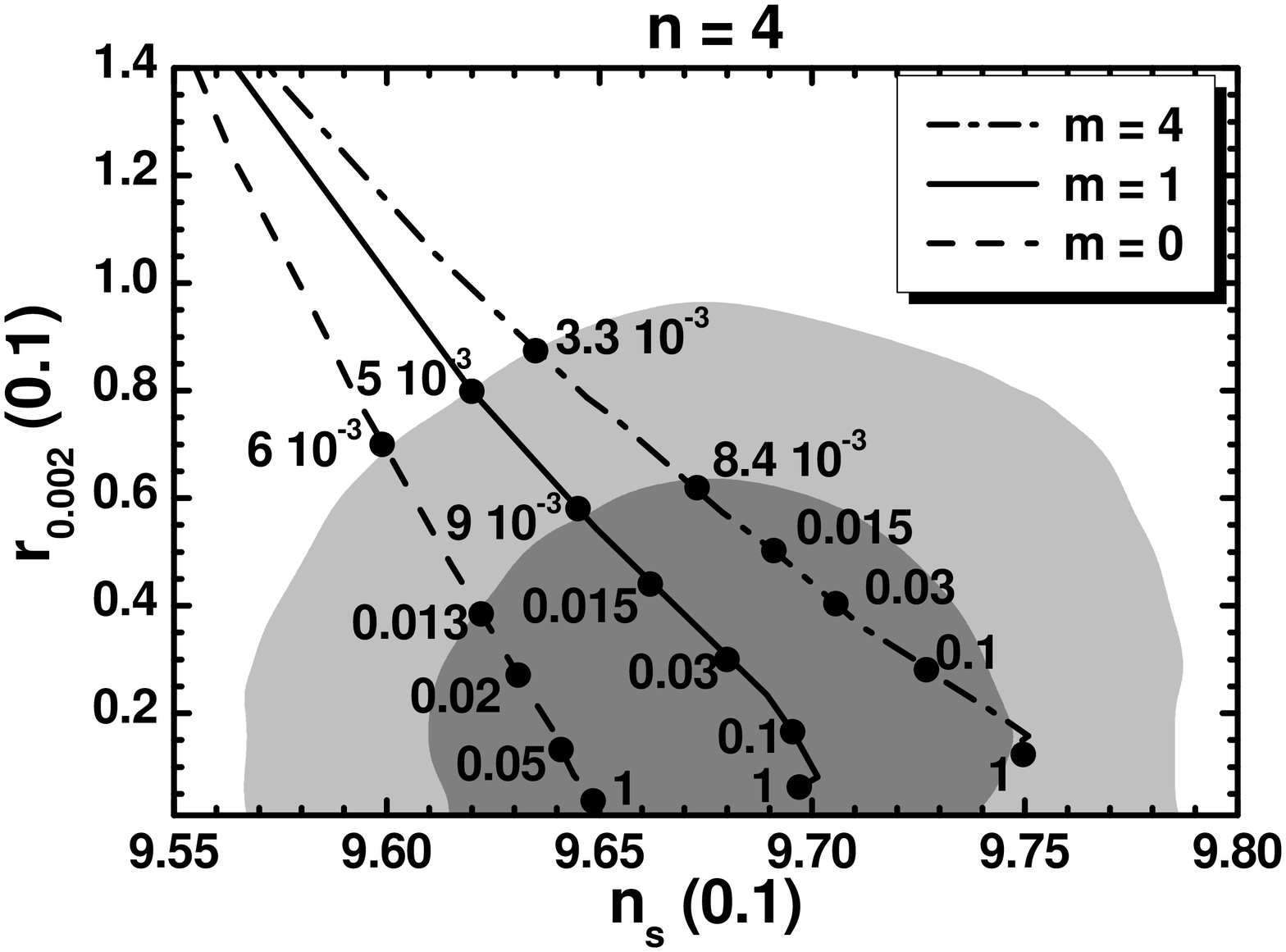,height=3.48in,angle=-90} \hfill
\end{minipage}\vspace*{-.2in}
\hfill \caption{\sl\small  Allowed curves in the $\ns-\rw$ plane
for $n=2$ and $4$, $m=0$ (dashed lines), $m=1$ (solid lines),
$m=4$ (dot-dashed lines), and various $\rs$'s indicated on the
curves. The marginalized joint $68\%$ [$95\%$] regions from \plk,
\bcp\ and BAO data are depicted by the dark [light] shaded
contours.}\label{fig1}
\end{figure}

From the plots in \Fref{fig1} we observe that, for low enough
$\rs$'s -- i.e. $\rs=10^{-4}$ and $0.001$ for $n=4$ and $2$ --,
the various lines converge to the $(\ns,\rw)$'s obtained within
the simplest models of chaotic inflation with the same $n$. At the
other end, the lines for $n=4$ terminate for $\rs=1$, beyond which
the theory ceases to be unitarity safe -- as anticipated in
\Sref{Qef} -- whereas the $n=2$ lines approach an attractor value,
comparable with the value in \Eref{nswmap1}, for any $m$.

For $m=0$ we reveal the results of \Sref{res0}, i.e. the displayed
lines are almost parallel for $\rw\geq0.02$ and converge at the
values in \Eref{nswmap1} -- for $n=4$ this is reached even for
$\rs=1$. Our estimations in Eqs.~(\ref{ns1}) -- (\ref{as1}) are in
agreement with the numerical results for $n=2$ and $\rs\lesssim1$
or $n=4$ and $\rs\lesssim0.05$. We observe that the $n=2$ line is
closer to the central values in \Eref{nswmap}  whereas the $n=4$
one deviates from those.

For $m>0$ the curves change slopes w.r.t to those with $m=0$ and
move to the right. As a consequence, for $n=4$ they span densely
the 1-$\sigma$ ranges in \Eref{nswmap} for quite natural $\rs$'s
-- e.g. $0.005\lesssim\rs\lesssim0.1$ for $m=1$. It is worth
mentioning that the requirement $\rs\leq1$ (for $n=4$) provides a
lower bound on $\rw$, which ranges from $0.004$ for $m=0$ to
$0.015$ (for $m=4$). Therefore, our results are testable in the
forthcoming experiments \cite{cmbpol} hunting for primordial
gravitational waves. Note, finally, that our findings in
Eqs.~(\ref{ns2}) -- (\ref{as2}) approximate fairly the numerical
outputs for $0.003\lesssim\rs\leq1$.


\section{Conclusions}\label{con}

We reviewed the implementation of kinetically modified nMI in both
a non-SUSY and a SUSY framework. The models are tied to the
potential $\Vjhi$ and the coupling function of the inflaton to
gravity given in \Eref{Vn} and the non-canonical kinetic mixing in
\Eref{fk}. This setting can be elegantly implemented in SUGRA too,
employing the super-{}and \Ka s given in Eqs.~(\ref{Wn}) and
(\ref{K1}) -- (\ref{K4}). Prominent in this realization is the
role of a shift-symmetric quadratic function $\hk$ in \Eref{hr}
which remains invisible in the SUGRA scalar potential while
dominates the canonical normalization of the inflaton. Using
$m\geq0$ and confining $\rs$ to the range $(3.3\cdot10^{-3}-1)$,
where the upper bound does not apply to the $n=2$ case, we
achieved observational predictions which may be tested in the near
future and converge towards the ``sweet'' spot of the present data
-- especially for $n=4$. These solutions can be attained even with
subplanckian values of the inflaton requiring large $\ck$'s and
without causing any problem with the perturbative unitarity. It is
gratifying, finally, that the most promising case of our proposal
with $n=4$ can be studied analytically and rather accurately.

\acknowledgments This research was supported from the MEC and
FEDER (EC) grants FPA2011-23596 and the Generalitat Valenciana
under grant PROMETEOII/2013/017.


\def\ijmp#1#2#3{{\emph{Int. Jour. Mod. Phys.}}
{\bf #1},~#3~(#2)}
\def\plb#1#2#3{{\emph{Phys. Lett.  B }}{\bf #1},~#3~(#2)}
\def\zpc#1#2#3{{Z. Phys. C }{\bf #1},~#3~(#2)}
\def\prl#1#2#3{{\emph{Phys. Rev. Lett.} }
{\bf #1},~#3~(#2)}
\def\rmp#1#2#3{{Rev. Mod. Phys.}
{\bf #1},~#3~(#2)}
\def\prep#1#2#3{\emph{Phys. Rep. }{\bf #1},~#3~(#2)}
\def\prd#1#2#3{{\emph{Phys. Rev.  D }}{\bf #1},~#3~(#2)}
\def\npb#1#2#3{{\emph{Nucl. Phys.} }{\bf B#1},~#3~(#2)}
\def\npps#1#2#3{{Nucl. Phys. B (Proc. Sup.)}
{\bf #1},~#3~(#2)}
\def\mpl#1#2#3{{Mod. Phys. Lett.}
{\bf #1},~#3~(#2)}
\def\arnps#1#2#3{{Annu. Rev. Nucl. Part. Sci.}
{\bf #1},~#3~(#2)}
\def\sjnp#1#2#3{{Sov. J. Nucl. Phys.}
{\bf #1},~#3~(#2)}
\def\jetp#1#2#3{{JETP Lett. }{\bf #1},~#3~(#2)}
\def\app#1#2#3{{Acta Phys. Polon.}
{\bf #1},~#3~(#2)}
\def\rnc#1#2#3{{Riv. Nuovo Cim.}
{\bf #1},~#3~(#2)}
\def\ap#1#2#3{{Ann. Phys. }{\bf #1},~#3~(#2)}
\def\ptp#1#2#3{{Prog. Theor. Phys.}
{\bf #1},~#3~(#2)}
\def\apjl#1#2#3{{Astrophys. J. Lett.}
{\bf #1},~#3~(#2)}
\def\n#1#2#3{{Nature }{\bf #1},~#3~(#2)}
\def\apj#1#2#3{{Astrophys. J.}
{\bf #1},~#3~(#2)}
\def\anj#1#2#3{{Astron. J. }{\bf #1},~#3~(#2)}
\def\mnras#1#2#3{{MNRAS }{\bf #1},~#3~(#2)}
\def\grg#1#2#3{{Gen. Rel. Grav.}
{\bf #1},~#3~(#2)}
\def\s#1#2#3{{Science }{\bf #1},~#3~(#2)}
\def\baas#1#2#3{{Bull. Am. Astron. Soc.}
{\bf #1},~#3~(#2)}
\def\ibid#1#2#3{{\it ibid. }{\bf #1},~#3~(#2)}
\def\cpc#1#2#3{{Comput. Phys. Commun.}
{\bf #1},~#3~(#2)}
\def\astp#1#2#3{{Astropart. Phys.}
{\bf #1},~#3~(#2)}
\def\epjc#1#2#3{{Eur. Phys. J. C}
{\bf #1},~#3~(#2)}
\def\nima#1#2#3{{Nucl. Instrum. Meth. A}
{\bf #1},~#3~(#2)}
\def\jhep#1#2#3{{\emph{JHEP} }
{\bf #1},~#3~(#2)}
\def\jcap#1#2#3{{\emph{JCAP} }
{\bf #1},~#3~(#2)}
\newcommand{\arxiv}[1]{{\ftn\tt  arXiv:#1}}


\begin{thebibliography}{99}

\bibitem{nMkin} C.~Pallis,~\prd{91}{2015}{123508} [\arxiv{1503.05887}].


\bibitem{old}  D. S. Salopek, J. R. Bond and J.M.
Bardeen, {\sl Phys. Rev. D }{\bf 40}, 1753 (1989);\\ F.L.~Bezrukov
and M.~Shaposhnikov, \plb{659}{2008}{703}  [\arxiv{0710.3755}].



\bibitem{nmi} C. Pallis, \plb{692}{2010}{287}
[\arxiv{1002.4765}];\\ C. Pallis and Q. Shafi,
\prd{86}{2012}{023523} [\arxiv{1204.0252}].

\bibitem{roest} R. Kallosh, A. Linde, and D. Roest,
{\sl Phys. Rev. Lett.} {\bf 112}, 011303 (2014)
[\arxiv{1310.3950}].

\bibitem{cutoff} J.L.F.~Barbon and J.R.~Espinosa,
\prd{79}{2009}{081302} [\arxiv{0903.0355}]; \\ C.P.~Burgess,
H.M.~Lee, and M.~Trott, \jhep{07}{2010}{007} [\arxiv{1002.2730}].

\bibitem{riotto} A.~Kehagias, A.M.~Dizgah, and A.~Riotto, \prd{89}{2014}{043527}
[\arxiv{1312.1155}].

\bibitem{plcp} \plk\ Collaboration, \arxiv{1502.02114}.


\bibitem{gws} P.A.R.~Ade {\it et al.}, \prl{114}{2015}{101301} [\arxiv{1502.00612}].



\bibitem{linde1} M.B.~Einhorn and D.R.T.~Jones,
\jhep{03}{2010}{026} [\arxiv{0912.2718}]; \\ H.M.~Lee,
\jcap{08}{2010}{003} [\arxiv{1005.2735}]; \\ S.~Ferrara \etal,
\prd{83}{2011}{025008} [\arxiv{1008.2942}]; \\ C.~Pallis and
N.~Toumbas, \jcap{02}{2011}{019} [\arxiv{1101.0325}].

\bibitem{nIG} C.~Pallis, \jcap{10}{2014}{058} [\arxiv{1407.8522}]; \\ C. Pallis and
Q.~Shafi, \jcap{03}{2015}{023} [\arxiv{1412.3757}].


\bibitem{lazarides}  G.~Lazarides and C.~Pallis,
\arxiv{1508.06682};\\ C. Pallis, to appear.

\bibitem{cmbpol} P.~Creminelli \etal, \arxiv{1502.01983}.



\end{thebibliography}
\end{document}